\documentclass[aps,prl,preprintnumbers,amsmath,amssymb,latexsym,array,enumerate,letter,twocolumn,superscriptaddress]{revtex4-1}
 \usepackage{titlesec}
   \titleformat{\section}[runin]
   {\normalfont\bfseries}{\quad\thesection.}{0.5em}{} [.]
  \titlespacing*{\section}{0pt}{0.2\baselineskip}{\baselineskip}

\usepackage{enumitem}
\usepackage[colorlinks=true,filecolor=red,citecolor=blue ]{hyperref}
\usepackage{mathtools}
\usepackage{subfigure}
\usepackage{graphicx}
\usepackage{dcolumn}
\usepackage{bm}
\usepackage{amssymb}
\usepackage{epsfig}
\usepackage{slashed}

\newcommand{\be}{\begin{equation}}
\newcommand{\ee}{\end{equation}}
\newcommand{\beq}{\begin{eqnarray}}
\newcommand{\eeq}{\end{eqnarray}}

\newcommand{\bpm}{\begin{pmatrix}}
\newcommand{\epm}{\end{pmatrix}}

\newcommand{\beqn}{\begin{eqnarray}}
\newcommand{\eeqn}{\end{eqnarray}}

\begin{document}

\title{Echoes of Inflationary First-Order Phase Transitions in the CMB}

\author{Hongliang Jiang}
\email{hjiangag@connect.ust.hk}
\affiliation{ Department of Physics, The Hong Kong University of Science and Technology, Clear Water Bay, Kowloon, Hong Kong S.A.R., P.R.C. }
\author{Tao Liu}
\email{taoliu@ust.hk}
 \affiliation { Department of Physics, The Hong Kong University of Science and Technology, Clear Water Bay, Kowloon, Hong Kong S.A.R., P.R.C. }
\author{Sichun Sun}
\email{sichun@uw.edu}
\affiliation{Jockey Club Institute for Advanced Study, The Hong Kong University of Science and Technology, Clear Water Bay, Kowloon, Hong Kong S.A.R., P.R.C. }
\author{Yi Wang}
\email{phyw@ust.hk}
\affiliation{ Department of Physics, The Hong Kong University of Science and Technology, Clear Water Bay, Kowloon, Hong Kong S.A.R., P.R.C. }

\begin{abstract}
Cosmological phase transitions (CPTs), such as the Grand Unified Theory (GUT) and the electroweak (EW) ones,  play a significant role in both particle physics and cosmology. In this letter, we propose to probe the first-order CPTs, by detecting gravitational waves (GWs) which are generated during the phase transitions through the cosmic microwave background (CMB). If happened around the inflation era, the first-order CPTs may yield low-frequency GWs due to bubble dynamics, leaving imprints on the CMB. In contrast to the nearly scale-invariant primordial GWs caused by vacuum fluctuation, these bubble-generated GWs are scale dependent and have non-trivial B-mode spectra. If decoupled from inflaton, the EWPT during inflation may serve as a probe for the one after reheating where the baryon asymmetry could be generated via EW baryogenesis (EWBG). The CMB thus provides a potential way to test the feasibility of the EWBG, complementary to the collider measurements of Higgs potential and the direct detection of GWs generated during EWPT.
 
\end{abstract}

\maketitle

\section{Introduction}

Phase transitions in particle physics have deep implications in cosmology. One famous example is the invention of inflation theory, which was originally motivated by addressing the missing magnetic  monopole  problem produced during the GUT phase transition \cite{Guth:1980zm}. Another example is related to cosmic baryon asymmetry. If the electroweak phase transition (EWPT) is of first order, the baryon asymmetry could be generated during the phase transition, with CP-violating Higgs couplings~\cite{Kuzmin:1985mm}.

With the discovery of the Higgs particle, the questions about the EWPT have  intensified. Though highly challenging, we expect to be able to probe the first-order EWPT by measuring Higgs self-interaction at High-Luminosity LHC or at future colliders (see, e.g.,~\cite{Noble:2007kk,Dolan:2012rv,Papaefstathiou:2012qe,Baglio:2012np,Liu:2014rva,Barr:2014sga,He:2015spf,Huang:2015tdv}). More generally, the cosmological phase transitions (CPTs) of first order are implemented via bubble nucleation. The expanding bubbles may collide with each other  or stir up turbulence and sound wave in the  thermal plasma, yielding gravitational waves (GWs) in spacetime~\cite{Witten:1984rs,Kamionkowski:1993fg,Hindmarsh:2013xza,Hindmarsh:2015qta}.  Particularly, if the phase transitions are of EW scale or PeV scale and happened after reheating~\cite{Kosowsky:1991ua, Ignatius:1993qn, Huber:2008hg,Cline:2006ts}, the produced GWs are characterized by a frequency $\gtrsim 10^{-4}$Hz \cite{Grojean:2006bp} that direct detection experiments, like Advanced LIGO \cite{Harry:2010zz}, Advanced Virgo \cite{Aasi:2013wya} and LISA \cite{AmaroSeoane:2012km}, are currently looking for or will look for. In this letter, we propose a new approach of probing the first-order CPTs,  by detecting the bubble-generated GWs through the cosmic microwave background (CMB).

The CMB temperature and polarization fluctuations provide us rich information about the primordial universe \cite{Ade:2015xua, Array:2015xqh}. Cosmic inflation \cite{Guth:1980zm} is the leading paradigm to seed those CMB fluctuations. The potential role of the CMB in probing the first-order CPTs was ignored. Because if the CPTs happened after reheating, the produced GWs have a characteristic frequency far beyond the scope of CMB.

One fact often ignored before about inflation is that an inflationary universe may undergo some thermal phase transition due to temperature decreasing, if the initial temperature of the universe is above the inflationary Hubble scale. In some physical contexts, such as the GUT and the recently proposed cosmological relaxation models aimed to solve the hierarchy problem~\cite{Graham:2015cka}, such phase transitions could be generic. These phase transitions, if of first order, may lead to entirely different cosmological consequences. Particularly, they are different from  old inflation scenarios, e.g.~\cite{Guth:1980zm,Turner:1992tz}, where inflation was driven by a first-order phase transition and thus the phase transition can not finish. In our scenarios, inflation occurred inside and outside the bubbles simultaneously, enabling the sub-horizon bubbles to collide with each other and hence generate GWs during inflation. Since the phase transitions happened during inflation, the produced GWs can be characterized by scales comparable to the size of the current universe with a scale dependent power spectrum. Thus they can be observed in the CMB potentially.

\begin{figure}
\includegraphics[width=7cm]{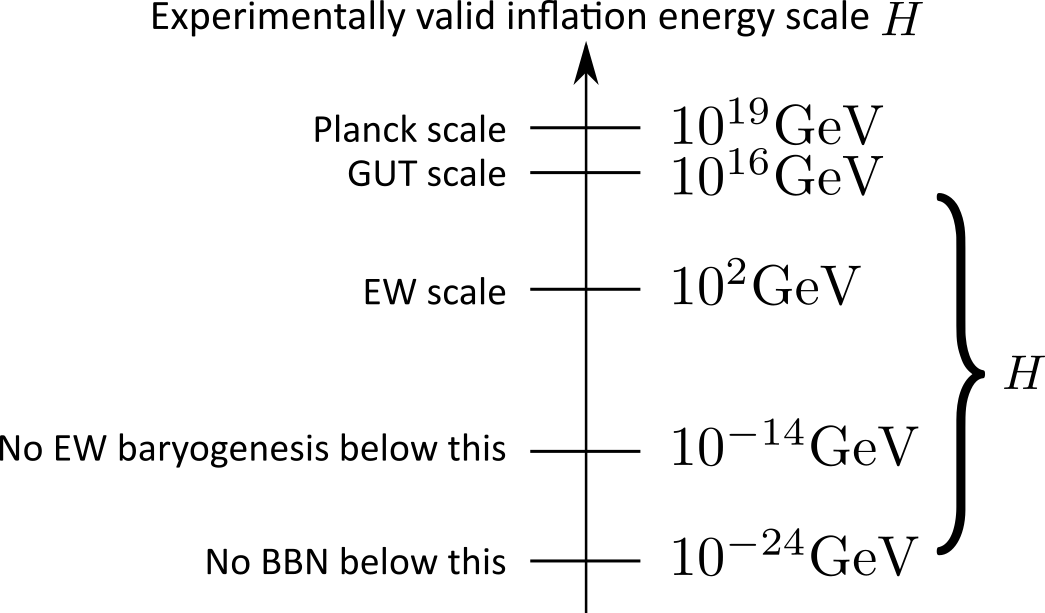}
\caption{ In various inflation models, Hubble constant during inflation can take different values from $10^{-24}$ GeV up to $10^{14}$GeV. The upper bound is set by the latest experiment \cite{Array:2015xqh}. When the Hubble constant is below $10^{-14}$ GeV, the universe can not reheat above $100$GeV later,  where the EW baryogenesis can be hardly achieved. Below $10^{-24}$ GeV, the reheated universe is too cool to have big bang nucleosynthesis (for a recent bound after \emph{Planck} 2015, see \cite{deSalas:2015glj}). }
\label{HubbleScale}
\end{figure}

\begin{figure}
\includegraphics[width=7cm]{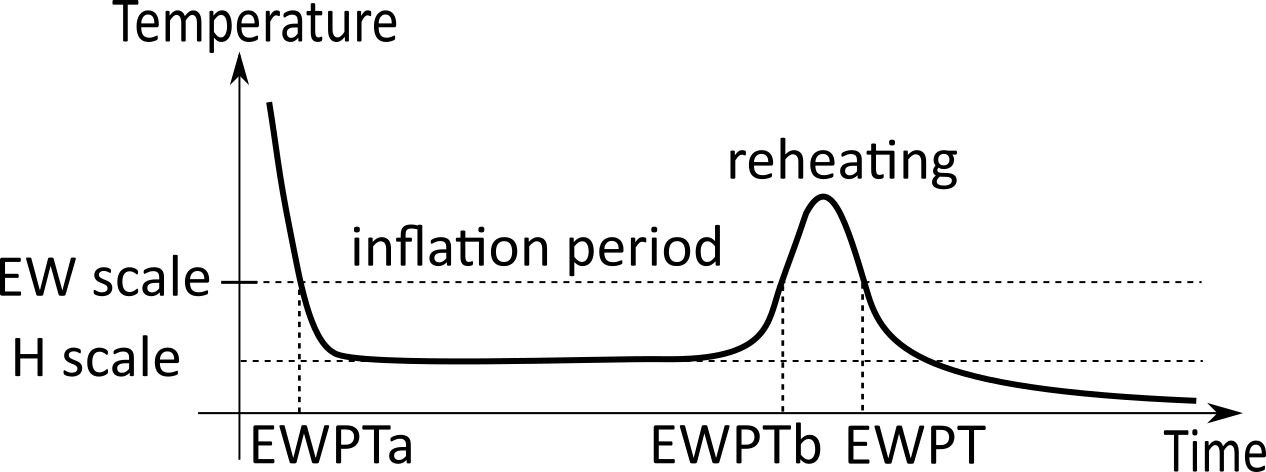}
\caption{ An exemplifying thermal temperature versus time relation in the early universe. After inflation reaches the attractor phase, the temperature comes mainly from the curvature contribution of the de-Sitter space, which is comparable with the Hubble constant. In addition to the EWPT after reheating, two more EWPTs are proposed, namely EWPTa during inflation, and EWPTb when the universe gets heated up during reheating. Here  $10^{-14}{\rm GeV}< H < \Lambda_{\rm EW}$ is assumed. EWPTa is the one relevant to discussions below.}
\label{temperatureEvolution}
\end{figure}

More explicitly, the temperature contribution from radiation drops quickly at the beginning of inflation, leaving only the Gibbons-Hawking temperature  $T_{GH}=H/(2\pi)$ \cite{Gibbons:1977mu}.  $T_{GH}$ can vary between $10^{14}$GeV to $10^{-24}$GeV (see Fig.~\ref{HubbleScale}). The vast span of the unknown energy scale  of inflation may encompass rich physics. Among different models of inflation with low energy scales~\cite{Dimopoulos:2004yb,Allahverdi:2006iq}: $H \sim T_{GH}  \leq 10^2$GeV, the EWPT will happen during the drastic decreasing of temperature at the beginning of inflation (denoted as ``EWPTa'' in Fig.~\ref{temperatureEvolution}). Note, for $H \sim 10^2 $GeV, the energy density is still as high as $\rho \sim (10^{10} \text{GeV})^4$, ensuring that inflation can happen. Moreover, the GUT phase transition, if exists, will happen in almost all the inflation scenarios.

\section{Gravitational wave spectra}\label{Sec_GW_Spectrum}

We derive the power spectrum $P_\gamma$ coming from scale dependent GWs produced in de-Sitter space here. During inflation, the action of GWs is
\begin{equation}
S=\frac{M_p^2}{8} \int  d\tau d^3 \bm x\; a^2\Big[(h_{ij}\rq{})^2-(\nabla h_{ij})^2 \Big]~,
\end{equation}
where the prime denotes the derivative with respect to conformal time $\tau$ and $a(\tau)\approx -1/H\tau$ is the scale factor. We introduce the polarization tensors $\epsilon^+_{ij}, \epsilon^\times_{ij}$ and decompose the gravitational fields:
\begin{align}
  h_{ij}(\bm{k}) = \frac{\sqrt2}{M_p} \left[ \gamma_+(\bm{k}) \epsilon^+_{ij}(\bm{k})
    + \gamma_\times (\bm{k}) \epsilon^\times_{ij}(\bm{k}) \right]~.
\end{align}
Then $\gamma_s$ can be quantized as
$
\gamma_s(\bm k,\tau)= v_k(\tau)a_{\bm k s}+v^*_k(\tau)a^\dagger_{-\bm k s }~,
$
where $a_{\bm k s}$ and  $a_{-\bm ks}^\dagger$ are  creation and annihilation operators.
Solving the equation of motion,  we can get the mode function $v_k$:
\begin{equation}
v_{k}(\tau)=\frac{H}{\sqrt{2k^3}}\Big(c_1( k) (1+ik\tau)e^{-ik\tau}+c_2( k) (1-ik\tau)e^{ik\tau} \Big) ~,
\end{equation}
where the coefficients $c_1(k),c_2(k)$ are subject to the consistency condition of quantization
$ |c_1|^2-|c_2|^2=1 $.

 The energy density of GWs is
\begin{equation}
 \rho_{GW}=\int \frac{ dk}{k} \frac{ k^3}{2\pi ^2} \Big(  |\dot{v}_{k} |^2 +\frac{k^2}{a^2}|v_{k} |^2  \Big)~.
\end{equation}
The gravitational energy spectrum is thus
\begin{eqnarray}
\Omega_{GW}(k,\tau)&=&k\frac{d\rho_{GW} }{dk}/ \rho_{\text{tot}} \\
&=&\frac{1 }{3H^2  M_p^2} \frac{ k^3}{2\pi ^2}
   \frac{|v_{k}\rq{}(\tau)|^2 +k^2|v_{k}(\tau)|^2 }{a(\tau)^2}~,
\end{eqnarray}
where we assume that the universe is spatially flat, meaning that $\rho_{\text{tot}}=\rho_{\text{critical}}=3H^2M_p^2$.  Particularly note that during inflation $\rho_{\text{tot}}=\rho_{\text{inflaton}}+\rho_{\text{rad}}+\rho_{\text{higgs}}$.

We can also calculate the power spectrum of GWs: 
\begin{equation}
P_\gamma(k,\tau)=\frac{4k^3}{\pi^2 M_p^2} |v_k(\tau)|^2 ~.
\end{equation}
This power spectrum of GWs contributes to both CMB temperature fluctuations and polarizations.

Both power spectrum and energy spectrum depend on the unknown functions $c_1(k),c_2(k)$. We are particularly  interested in the power spectrum at the time $\tau_{\text{obs}}\rightarrow 0$ when the modes exit the horizon and do not evolve anymore. While in general the GWs generated at time $\tau_*$ may be  either sub-horizon or super-horizon. The relation between them yields:
\begin{equation}
P_\gamma( \tau_{\text{obs}} )=24H^2 \Big( \frac{a(\tau_*)}{k} \Big)^2
 \frac{k^2|v_k(\tau_{\text{obs}})|^2}{k^2|v_k(\tau_*)|^2+|v_k\rq(\tau_*)|^2} \Omega_{GW}( \tau_*)~.
\end{equation}

We consider the classical limit $c_1\approx  c_2 \gg 1$.  Inserting the mode functions, we can get the relations for sub-horizon and super-horizon modes respectively,
 \begin{equation}\label{P_and_Omega}
 P_{\gamma}(k,\tau_{\text{obs}}\rightarrow0)=
\begin{cases}
  24 \Big( \frac{a(\tau_*)H}{k} \Big)^4   \Omega_{GW}(k,\tau_*) ,  & |k\tau_*| \gg 1
\\
 24 \Big( \frac{a(\tau_*)H}{k} \Big)^2   \Omega_{GW}(k,\tau_*) ,   & |k\tau_*| \ll 1
\end{cases} ~.
 \end{equation}
We will mainly discuss the GWs generated by sub-horizon bubbles below and leave the super-horizon case to the final discussion.

\section{Gravitational waves by the bubbles}\label{Sec_Bubble_GW_Spectrum}
The sub-horizon case in our inflationary scenarios is similar to the previous semi-numerical studies on the EW phase transitions such as \cite{Huber:2008hg,Kamionkowski:1993fg}. Usually the phase transition is a rapid process compared to the Hubble time and thus the effect of expansion of the universe can be ignored even during inflation. The only difference is that the total energy density $\rho_{\text{tot}}$ is higher with the contribution from the inflaton to drive inflation. Sub-horizon bubbles can stir up turbulence~\cite{Kamionkowski:1993fg} and sound wave~\cite{Hindmarsh:2013xza,Hindmarsh:2015qta} in thermal plasma and thus generate GWs.  For illustration, however, we focus on the GWs generated via bubble collisions, and apply ``envelope approximation''  to bubble walls only, with thermal effects neglected (similar to the physics of ``runaway bubbles in vacuum'' . For a review, see, e.g.,~\cite{Caprini:2015zlo}). It is straightforward to generalize the discussions to the GWs caused by sound wave and turbulence in thermal transitions.

Based on the similarity between inflationary and late phase transitions, one immediately obtains original energy spectrum~\cite{Kamionkowski:1993fg,Huber:2008hg,Weir:2016tov}
\begin{eqnarray}
    \Omega_{GW}(k)&=&\Omega_{GW}^{\text{crit}} \frac{(a+b) k_{\text{crit}} ^b k^a}{ b k_{\text{crit}} ^{a+b} +a k^{a+b} }~,
     \\
    \Omega_{GW}^{\text{crit}} &=& \frac{0.11 v_b^3 }{0.42+v_b^2} \kappa^2 \Big(\frac{H}{\beta}\Big)^2
   \Big( \frac{\rho_{\text{higgs}} }{\rho_{\text{tot}}}\Big)^2 ~,
    \end{eqnarray}
where $a,b$ are exponents parameterizing the scale dependence of the spectrum. Notice that for usual late phase transition we have $\rho_{\text{tot}} \rightarrow  \rho_{\text{higgs}}+\rho_{\text{rad}}$, where we can recover the standard results in the literature.

The critical point $k_{\text{crit}}$ is also the peak momentum of energy spectrum, which is given by
$
 k_{\text{crit}}/[2\pi a(\tau_* )\beta ] = 0.62/(1.8-0.1v_b+v_b^2).
$
Here a relativistic expansion rate $v_b\approx 1$ is typically expected, yielding $a\approx 2.8$, $b\approx 1$,  as discussed in~\cite{Huber:2008hg}. The parameter $\beta^{-1}$ is approximately the duration of the phase transition. In sub-horizon case we have $H/\beta <1$ although the specific values strongly depend on the shapes of the scalar potentials~\cite{Kehayias:2009tn}.  $\kappa$ denotes the  efficiency of converting vacuum energy into the bubble wall kinetic energy instead of thermal energy~\cite{Kamionkowski:1993fg,Huber:2008hg}. It depends on $\alpha= \rho_{\text{higgs}}/\rho_{\text{rad}}$, the ratio between the Higgs vacuum energy density and the radiation energy density, as well as the bubble nucleation rate.

We can then finally determine the power spectrum of GW generated by bubbles. Using Eq.~(\ref{P_and_Omega}), we arrive at:
\begin{equation}\label{final_powerS}
P_{\gamma}(k,\tau_{\text{obs}}\rightarrow 0 )=P_{\gamma}^{\text{crit}}  \Big( \frac{k_{\text{crit}}}{k}\Big)^4\frac{(a+b) k_{\text{crit}} ^b k^a}{ b k_{\text{crit}} ^{a+b} +a k^{a+b} }~,  
\end{equation}
where $P_{\gamma}^{\text{crit}}$ is the power spectrum at the critical point:
$
P_{\gamma}^{\text{crit}} =24   \Big(\frac{a(\tau_*)H}{k_{\text{crit}}}\Big)^4  \Omega_{GW}^{\text{crit}}~.
$
An estimation yields: $P_{\gamma}^{\text{crit}}\sim\Big(\frac{H}{\beta}\Big)^6
   \Big( \frac{\rho_{\text{higgs}} }{\rho_{\text{tot}}}\Big)^2$ for the sub-horizon case. The power spectrum is scale dependent and provides us a way to probe phase transition parameters.

\begin{figure}[t ]
\includegraphics[width=0.485\textwidth]{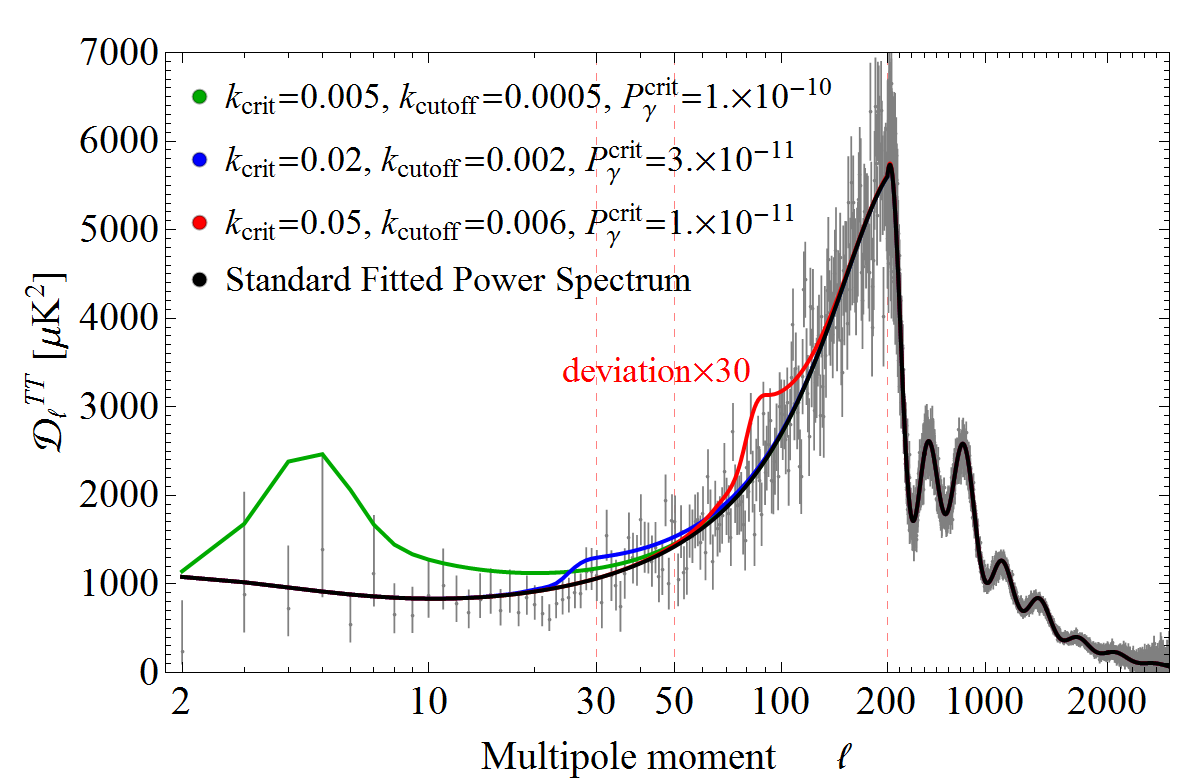}
\caption{CMB temperature power spectrum. The gray points and error bars are from \emph{Planck} 2015 while the black curve is the best fit of \emph{Planck} 2015. The green, blue and red curves represent power spectra (including the GWs) in three model-independent benchmarks, which could be projected to various scenarios in particle physics.  Note that for the red curve,  its deviation from the standard one is magnified by 30 times (The unit of $k$ is Mpc$^{-1}$ here and below.).}
\label{CMB_TT}
\end{figure}

\begin{figure} [t]
\includegraphics[width=0.485\textwidth]{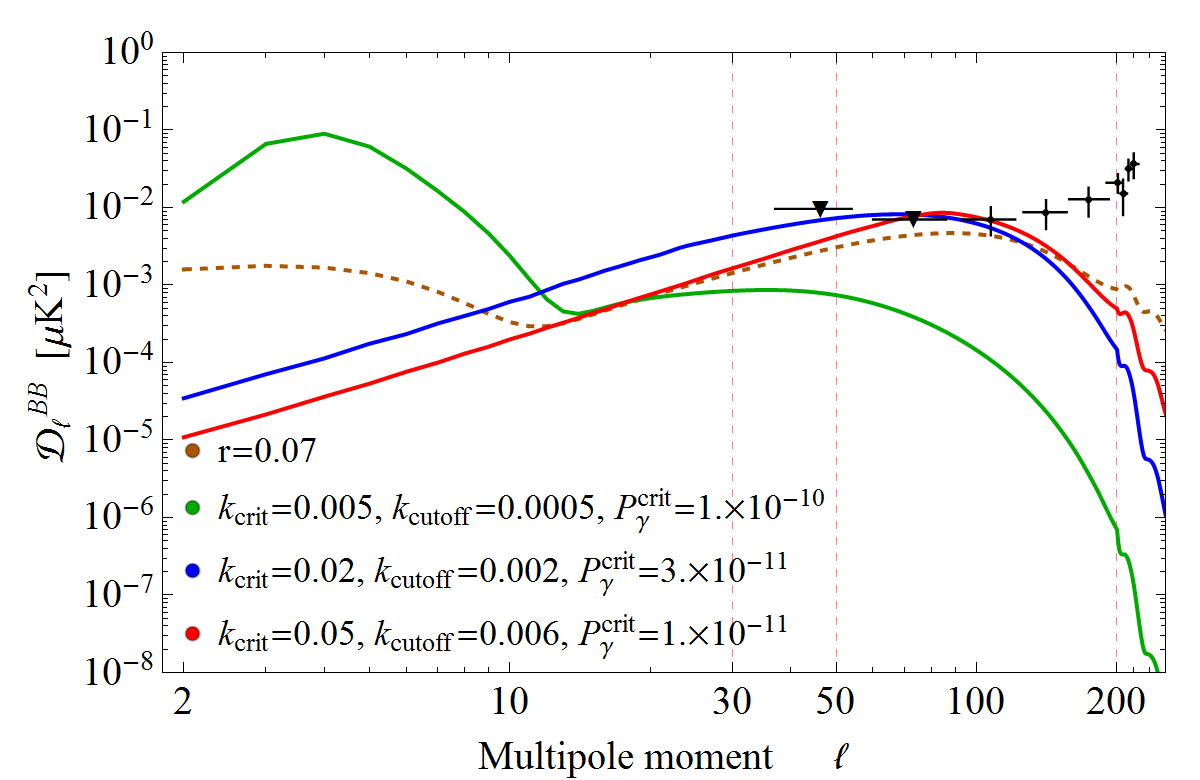}
\caption{B-mode power spectrum from phase transition bubbles.  For comparison, the primordial GWs from quantum vacuum fluctuations are showed in dashed line with tensor-to-scalar ratio $r=0.07$. The black symbols at the right-upper corner represent the CMB component bandpowers obtained from BICEP2 \& {\it Keck  Array}  experiments, with error bars denoting 68\%  credible intervals and downward triangles indicating the 95\% upper bound~\cite{Array:2015xqh}. The colored solid lines correspond to the benchmarks defined in Fig.~\ref{CMB_TT}. The two dotted lines denoted effective noise level of two representative experiments POLARBEAR~\cite{Ade:2014afa} and SPIDER~\cite{Crill:2008rd}, respectively, based on a Fisher forecast analysis as done in~\cite{Ma:2010yb,Wang:2014oka,Creminelli:2015oda}.}
\label{CMB_Bmode}
\end{figure}

\begin{figure} [t]
\includegraphics[width=0.485\textwidth]{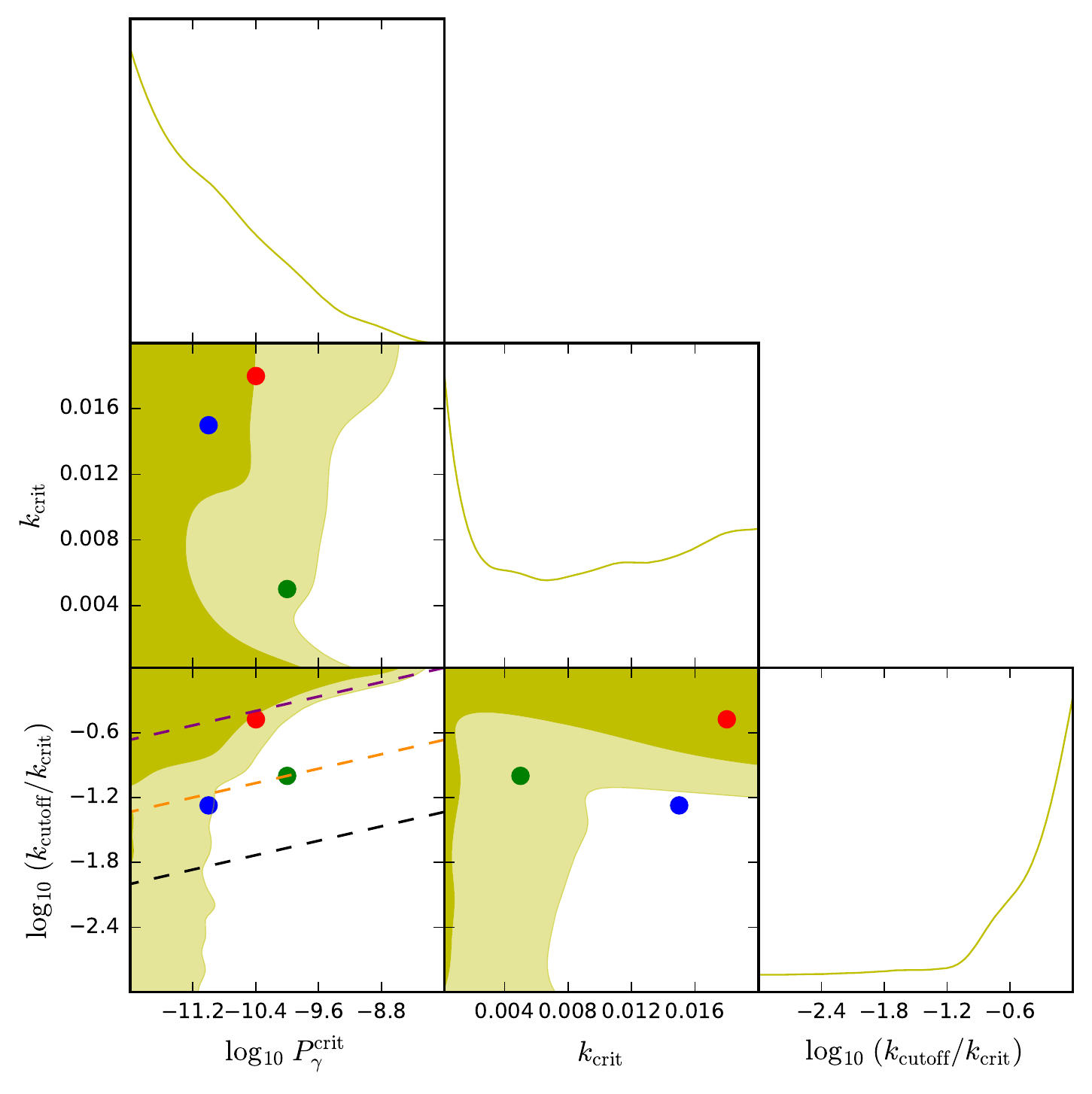}
\caption{Constraints of \emph{Planck}$ 2015+$BICEP2/Keck data for $\{P_{\gamma}^{\text{crit}}$,  $k_{\text{crit}}$, $k_{\text{cutoff}}/k_{\text{crit}}\}$. The one-parameter panels show the parameter likelihood. In the two-parameter panels, dark yellow and yellow are the marginalized 1$\sigma$ and 2$\sigma$ contours, respectively. The colored points correspond to the benchmarks defined in Fig.~\ref{CMB_TT}. The black, orange, purple dashed lines in the left-bottom panel corresponds to $ \Big ( \frac{\rho_{\rm higgs}}{\rho_{\rm tot}} \Big)^2 =1, 10^{-4}$ and $10^{-8}$, respectively.} 
\label{param_constrain}
\end{figure}


 \section{Imprints in the CMB}\label{Sec_CMBimprint}

The CMB spectrum can be obtained by inputting the power spectrum into the \textsf{CLASS} \cite{Blas} where the transfer function is calculated. As we see from Eq.~(\ref{final_powerS}), the power spectrum diverges as $k\rightarrow 0$. This divergence is unphysical as the GW generating formulae break down at super horizon scale. 
We can introduce the horizon scale as a natural cut-off. The power spectrum $P_{\gamma}(k,\tau_{\text{obs}}\rightarrow 0 )$ of the GW generated by bubbles therefore are described by three free parameters: $P_{\gamma}^{\text{crit}}$,  $k_{\text{crit}}$, $k_{\text{cutoff}}$. The physical $k_{\text{cutoff-physical}}$ is given by $k_{\text{cutoff-physical}}=k_{\text{cutoff}}/a_* \sim H $, and the physical critical momentum is related to the bubble size via $k_{\text{crit} }/a_*\sim R_b^{-1}$. The comoving momenta are
\begin{equation}
k_{\text{crit}} \sim \frac{1}{v_b}   \frac{\beta}{H}    e^{N_*}     k_0, \quad
k_{\text{cutoff}} \sim   e^{N_*}     k_0,
\end{equation}
yielding $ k_{\text{cutoff}}/k_{\text{crit}}  \sim H/\beta$. Here $N_*$ is the e-folds of phase transition counting from the time that the largest mode $k_0$ exits the horizon.  We choose the scale factor  today $a_0$ as one, thus the largest physical mode today is $k_0=0.0002 \text{Mpc}^{-1}$ as the inverse of observed universe size. Approximately we can find $k_0 $ corresponding to the position at CMB multipole $\ell_0\sim 2$. Then we arrive at relations: $\ell_{\text{crit}} \sim 2e^{N_*}\beta/(v_b H  )   ,\ell_{\text{cutoff}} \sim  2 e^{N_*}$.

We plot the CMB spectrum with the GW contributions: $P_{\gamma}^{\text{crit}}\sim    10^{-10} -10^{-11}$ in Fig.~\ref{CMB_TT} and Fig.~\ref{CMB_Bmode}. The phase transition happened at early time of inflation era generates a new peak on CMB temperature spectrum. The generic peak positions are at $\ell<200$. For $\ell>200$, GW modes have already entered the horizon at recombination and thus are subject to rapid decay, posing a greater challenge in experiments. This also implies that the observable phase transitions should happen soon after the largest mode exits horizon with $N_*\lesssim 5$. The amplitudes of those peaks encode the information about the energy scale of inflation and the phase transitions.

Fig.~\ref{CMB_Bmode} shows the tensorial B-mode spectra, which have not been observed  yet in experiment.  For primordial GWs caused by vacuum fluctuations, B-mode spectrums are known to have the recombination  peak at $\ell \sim 100$ and the reionization bump  at $\ell< 10$.  For $10 \lesssim \ell \lesssim 100 $, the spectrum roughly scales like $\ell^2$ due to the dominant contribution from recombination~\cite{Pritchard:2004qp}.
 For our scenarios, the power spectrums   roughly scale like $k^{-1}$ when $k_{\text{cutoff}}<k<k_{\text{crit}}$ and like $k^{-5}$ when $k>k_{\text{crit}}$.  Therefore, if   $10 \lesssim \ell_{\text{cutoff}} \lesssim 100 $ and $\ell_{\text{crit}}\gtrsim 100$, B-mode spectrum will scale like $\ell$ for $\ell<100$. Thus we expect a peak to appear near the recombination peak. For small enough $\ell_{\text{cutoff}} \lesssim 10$, the reionization bump and the primordial GW peak in our scenario may lead to more complicated multiple dependence, e.g. as  the blue curves shown in  Fig.~\ref{CMB_Bmode}. The ongoing and future CMB experiments are expected to extend the BICEP2 sensitivities to large angular scales. In Fig.~\ref{CMB_Bmode} we also show the effective noise level of two representative experiments POLARBEAR~\cite{Ade:2014afa} and SPIDER~\cite{Crill:2008rd}, respectively, based on a Fisher forecast analysis as done in~\cite{Ma:2010yb,Wang:2014oka,Creminelli:2015oda}. The effective noise includes contributions from instrumental noise, residual foreground contamination, and the gravitational lensing (without delensing). The dash-dotted part ($l<20$) of the POLARBEA curve represents its limitation in probing large angular scales due to its relatively small survey areas in sky as a ground-based experiment.

In Fig.~\ref{param_constrain}, we show the constraints of the \emph{Planck}2015 $+$BICEP2/Keck data for the three parameters $\{P_{\gamma}^{\text{crit}}$, $k_{\text{crit}}$, $k_{\text{cutoff}}/k_{\text{crit}}\}$, based on a Bayesian analysis using \textsf{CosmoMC} \cite{Lewis:2002ah}. The constraints are more sensitive to $P_{\gamma}^{\text{crit}}$ and  $k_{\text{cutoff}}/k_{\text{crit}}$, compared to $k_{\rm crit}$, as they characterize the overall magnitude of the GWs power spectrum (see Eq.~(\ref{final_powerS})). In the two-parameter panels, the three benchmark scenarios defined in Fig.~\ref{CMB_TT} are also marked with colored points: green, blue and red. At $2\sigma$ C.L., the green one has been excluded, whereas the blue and red ones are marginally allowed and safe, respectively. In the left-bottom panel, the region below the black dashed line is theoretically forbidden, because of the requirement 
\begin{eqnarray}
P_{\gamma}^{\text{crit}}\sim\Big(\frac{H}{\beta}\Big)^6
   \Big( \frac{\rho_{\text{higgs}} }{\rho_{\text{tot}}}\Big)^2   \lesssim \Big(\frac{H}{\beta}\Big)^6  \sim \Big(\frac{k_{\text{cutoff}}}{k_{\text{crit}}} \Big)^6 \ .
\end{eqnarray}
The lower bound for $P_{\gamma}^{\text{crit}}$ is model-dependent. Below are two concrete examples in particle physics:


\begin{itemize}[noitemsep,topsep=0pt]

 \item GUT scenarios: The GUT phase transition happens around $10^{16}$ GeV \cite{Georgi:1974yf}, which can be of strong first-order in some generic scenarios~\cite{Guth:1979bh}.  To avoid re-introducing the problem of magnetic monopoles, the GUT phase transition can only happen before or soon after the begin of inflation, with $\frac{\Lambda_{\rm GUT}^2}{M_p} \lesssim H \lesssim 10^{14} {\rm GeV} < \Lambda_{\rm GUT}$. Thus the projection of the allowed parameter space in the $\log_{10}P_{\gamma}^{\text{crit}} - \log_{10}\frac{k_{\text{cutoff}}}{k_{\text{crit}}}$ plane is along the black-dashed line (see Fig.~\ref{param_constrain}), yielding  $P_{\gamma}^{\text{crit}}\sim \Big(\frac{H}{\beta}\Big)^6 $.   The inflation is of high scale here, potentially generating detectable scale-invariant GWs via vacuum fluctuation. If both the vacuum and bubble GWs are detected, we may infer that the PT is at GUT scale instead of EW scale.

 \item EW scenarios: A first-order EWPT can be achieved in various theories, e.g. \cite{Huet:1995sh,Trodden:1998ym,Morrissey:2012db,Cohen:1993nk, Apreda:2001us,Kang:2004pp}. Unlike what happens to the GUT scenario, the inflation needs to be low-scale here. The theoretically allowed range for $P_{\gamma}^{\text{crit}}$ is broad, say, $\sim \mathcal O(10^{-60}-1) \Big(\frac{k_{\text{cutoff}}}{k_{\text{crit}}}\Big)^6$, extending from the black-dashed line to above (see Fig.~\ref{param_constrain}). If taking $(H/\beta)^6 \sim 10^{-6}, \rho_\text{higgs} \sim (10^3\text{GeV})^4$, and $H \sim 10^{-11} \text{GeV} $, then we have  $P_{\gamma}^{\text{crit}}\sim 10^{-10}$. This possibility has been excluded, if $N_* \lesssim 5$ or $k_{\rm crit}< 0.3$, as is indicated by the benchmark point in green in Fig.~\ref{param_constrain}.  The baryon asymmetry in the universe today can not be directly connected to the first-order EWPT during inflation, due to inflationary dilution. However, if decoupled from inflaton, the effective Higgs potential is sensitive to the temperature of thermal plasma only, subjecting to a negligible curvature correction of order $\mathcal {O}\Big(\frac{H}{\Lambda_{\rm EW}}\Big)$. The EWPT during inflation (``EWPTa'' in Fig.~\ref{temperatureEvolution}) thus may serve as a probe for the one after reheating and hence to test the feasibility of the EWBG.

\end{itemize}

\section{Discussions}\label{Sec_conclusion}
In this letter we propose an indirect approach to probe the first-order CPTs, such as the GUT  phase transition and the EWPT, by detecting the GWs through the CMB. These GWs are generated during inflation via  bubble collisions or by the turbulence caused by bubble motion in the thermal plasma. Generically different from the primordial GWs caused by vacuum fluctuation during inflation, which have been extensively studied in the past decades, the bubble-generated GWs are scale dependent, potentially yielding non-trivial temperature and B-mode spectra in the CMB. Therefore, these GWs (or the first-order CPTs under exploration) represent a new class of physics targets, characterized by a scale-dependent power spectrum, for the ongoing and future CMB experiments to explore. 

The large-scale scalar power spectrum caused by the first-order CPTs during inflation might be suppressed. This is due to the relatively large values of the slow-roll parameter when the thermal radiation is diluted. The further discussion on the density fluctuation depends on the details of the specific inflation models. The GWs and density fluctuations generated by super-horizon bubbles share some features in physics discussed above, which may leave imprints on the CMB as well. It remains interesting to investigate reheating in more details, especially in the EW secnarios. We expect preheating  \cite{Traschen:1990sw,Kofman:1997yn}  may provide efficient reheating, if inflaton and Higgs fields are decoupled. We leave the detailed study to a future work.

\begin{acknowledgments}
\noindent\textbf{Acknowledgments. }The authors would like to thank Andrew Cohen, Lam Hui, Yin-Zhe Ma, Ann E. Nelson, Henry Tye and Lian-Tao Wang for useful discussions. This work was supported by the CRF Grants of the Government of the Hong Kong SAR under HKUST4/CRF/13G.
\end{acknowledgments}

\end{document}